\documentclass{iopart}
\usepackage{iopams}
\usepackage{graphicx}
\usepackage{dcolumn}
\usepackage{bm}
\usepackage{xspace}
\usepackage{dsfont}
\usepackage{psfrag}
\usepackage{bbm}
\usepackage{hyperref}  
\usepackage{float}
\usepackage{subfig}
\usepackage{textcomp}
\usepackage{ulem}
\usepackage{color}

\begin{document}

\title{An Efficient Integrated Two-Color Source for Heralded Single Photons}
\author{S. Krapick, H. Herrmann, V. Quiring, B. Brecht, H. Suche and Ch. Silberhorn}
\address{University of Paderborn, Department of Physics, Integrated Quantum Optics Group, Warburger Str. 100, 33098 Paderborn, Germany}
\eads{krapick@mail.upb.de}

\begin{abstract}

We present a pulsed and integrated, highly non-degenerate parametric downconversion (PDC) source of heralded single photons at telecom wavelengths, paired with heralding photons around 800nm. The active PDC section is combined with a passive, integrated wavelength division demultiplexer on-chip, which allows for the spatial separation of signal and idler photons with efficiencies of more than 96.5\%, as well as with multi-band reflection and anti-reflection coatings which facilitate low incoupling losses and a pump suppression at the output of the device of more than 99\%. Our device is capable of preparing single photons with efficiencies of 60\% with a coincidences-to-accidentals ratio
exceeding 7400. Likewise it shows practically no significant background noise compared to continuous wave realizations. For low pump powers, we measure a conditioned second-order correlation function of $g^{(2)}=3.8\cdot 10^{-3}$, which proves almost pure single photon generation. In addition, our source can feature a high brightness of $\langle n_{\mathrm{pulse}}\rangle=0.24$ generated photon pairs per pump pulse at pump power levels below $100\,\mu$W. The high quality of the pulsed PDC process in conjunction with the integration of highly efficient passive elements makes our device a promising candidate for future quantum networking applications, where an efficient miniaturization plays a crucial role.
\end{abstract}


\maketitle

\section{Introduction}
\label{Sec1}

Recent progress in the field of quantum information processing has highlighted the prospects of using integrated optic devices for quantum applications \cite{Tanzilli2012,Martin2012a,Christ2012c,Sangouard2011,Bonneau2012}. Integrated quantum photonics offers several advantages in comparison to free-space experimental setups with bulk optic components \cite{Pomarico2012}. The miniaturization of systems with increased complexity does not only drastically reduce the required space and paves the way for future commercialization, but it also enables the implementation of optical networks with a large number of optical modes and extremely high stability.

In $2008$ Politi and co-workers \cite{Politi2008} have demonstrated the first quantum interference and photonic gates on-chip, while different groups have developed sophisticated and integrated experiments with two-photon interference \cite{Marshall2009,Peruzzo2011} or photon-entanglement \cite{Matthews2009,Shadbolt2012}, controlled qubit-operations \cite{Li2011} as well as controlled phase shifts in linear optical circuits \cite{Bonneau2012,Smith2009}. In $2012$ Metcalf et al have realized the first three-photon experiment inside a linear optical network \cite{Metcalf2012} and recent research on Boson sampling in an integrated device demonstrates four-photon quantum interference\cite{Spring2012}.

However, in all of these experiments the preparation of the photon pairs has actually been performed outside the integrated devices employing traditional bulk crystal parametric down-conversion sources. The efficient coupling between these sources and the integrated circuit remains one bottleneck for designing systems with increasing complexity.

On the other hand remarkable efforts have been devoted to the development of integrated PDC sources for photon pair generation inside channel waveguides \cite{Tanzilli2001,Banaszek2001,Booth2002,Chen2009,Bonneau2012a,Kaiser2012,Karpinski2012,Solntsev2012} over the last decades. The main benefits of guided-wave PDC processes include high conversion efficiencies and spatial mode control.
 
Among several fabrication techniques with different nonlinear materials used for the implementation of PDC sources (see for example \cite{Banaszek2001,Pomarico2009,Alibart2005,Martin2010,Fiorentino2007}), titanium-indiffusion is a standard method to manufacture waveguides in lithium niobate (Ti:LN). It provides extremely low loss \cite{Regener1985} and the guiding of both polarizations.

In this paper we report on the fabrication and analysis of an integrated type-I PDC source in titanium-indiffused periodically poled Z-cut lithium niobate waveguides (Ti:PPLN) for heralded single photons. We combine an optically active component providing parametric down-conversion of high brightness with a passive element on-chip allowing for excellent spatial separation of the converted, highly non-degenerate photon pairs. This enables us, on one hand, to transmit prepared PDC states via low-loss fibers at the telecom L-band around $1575$ nm and on the other hand to herald these states with off-the-shelf silicon-based detectors at around $800$ nm. Together with two home-deposited endface coatings for close-to-perfect incoupling and outcoupling of the respective wavelengths, our device is a potential candidate to serve as a basic building block in more complex and miniaturized quantum network realizations in the future.

\section{Device design, characterization method and technology}
\label{Sec2}

\begin{figure}
\includegraphics[width=0.99\linewidth]{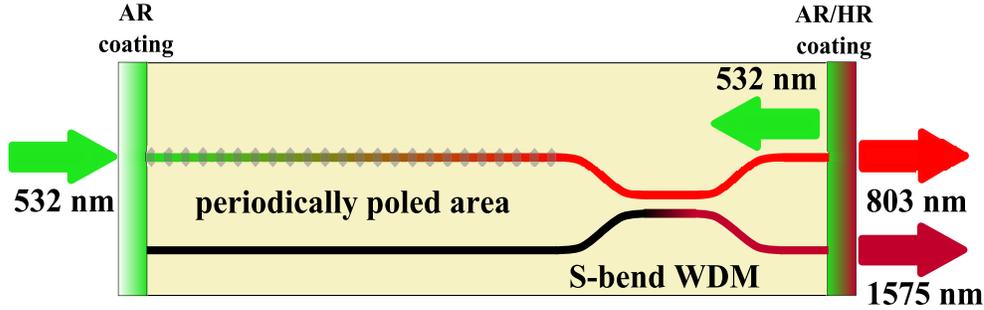}
\caption{\label{Pic1} Schematic of the Ti:PPLN photon pair source}
\end{figure}

Our PDC source basically consists of two integrated building blocks. \Fref {Pic1} shows the general design of the whole device. In the periodically poled area (first building block) the PDC process takes place and the integrated S-bend type wavelength division demultiplexer (WDM coupler, second building block) allows for the subsequent spatial separation of signal and idler photons. Home-deposited high quality optical endface coatings provide our PDC source with low coupling losses of the respective colors.

\subsection{First building block - PDC process}
\label{Sec21}

In $\chi^{(2)}$-nonlinear lithium niobate, parametric down-conversion represents a three-wave-mixing process, where one pump photon (p) decays into two daughter photons, commonly labeled signal (s) and idler (i) with the signal photon having the higher angular frequency. For this process, the conservation of photon energy
\begin{equation}
\label{1}
\hbar\omega_{\mathrm{p}}=\hbar\omega_{\mathrm{s}}+\hbar\omega_{\mathrm{i}}
\end{equation}
must be fulfilled.

In waveguides, the momentum conservation of the collinear travelling waves can be represented by the corresponding effective refractive indices of the individual guided modes of pump, signal and idler beams. For the ferroelectric lithium niobate an occuring phase mismatch
\begin{equation}
\label{4}
\Delta k=k\left( \lambda_{\mathrm{p}} \right)-k\left( \lambda_{\mathrm{s}} \right)-k\left( \lambda_{\mathrm{i}} \right),
\end{equation}
caused by dispersion, can be compensated by periodical inversion of the spontaneous material polarization, such that first-order quasi-phasematching with
\begin{equation}
\label{5}
\Delta k-\frac{2\pi}{\Lambda_{\mathrm{G}}}=0,
\end{equation}
for the appropriate period $\Lambda_{\mathrm{G}}$ of the inverted domains is achieved. The technology to achieve periodic poling is explained in \sref{Sec24} in more detail.

Quasi-phasematching allows for arbitrary wavelength combinations by choosing appropriate poling periods. However, short pump wavelengths in the visible regime require small poling periods. For fixed poling periods and pump wavelengths, the wavelength combination of signal and idler is determined. In this case, wavelength tuning of the signal and idler wavelength is only possible thermally due to the temperature dependence of the effective  refractive indices.

Here we focus on a type-I PDC process, in which pump photons at $532$ nm wavelength decay into non-degenerate signal and idler photons of TM polarization, accessing the highest nonlinear tensor element in lithium niobate $d_{33}$ for high conversion efficiencies. In order to determine the degree of non-degeneracy we varied the poling period of the device around $6.8\,\mu$m in several on-chip test structures.

\subsection{Heralding and conditioned $g^{(2)}(0)$}
\label{Sec22}

In $1977$ Kimble et al \cite{Kimble1977} demonstrated for the first time photon anti-bunching on sodium atomic emission. A well established way to estimate the mean photon number of a PDC source is the anti-correlation parameter $\alpha$. In 1986 Grangier et al \cite{Grangier1986} applied the concept of Clauser \cite{Clauser1974} in a Hanbury Brown-Twiss geometry \cite{HanburyBrown1956} to characterize anti-correlation phenomena by conditioned preparation of single photon states. The value of $\alpha$ can be determined by relating the count rates of binary photon detectors \cite{URen2005} when performing conditioned correlation measurements.

In our experiments we pump our PDC source with pulsed light. For low pump powers one pair of photons is generated with a probability much lower than unity and higher photon number contributions are insignificant. When increasing the pump power, higher pair generation rates are possible, however at the cost of increasing higher order components. These are detrimental for the quality of the heralded single photons, such that appearing higher order photon pairs should be carefully quantified.

Note that in the following we do not distinguish between the terms $trigger$, $signal\;photons$ and $heralding\; photons$, anymore.

\begin{figure}
\includegraphics[width=0.49\linewidth]{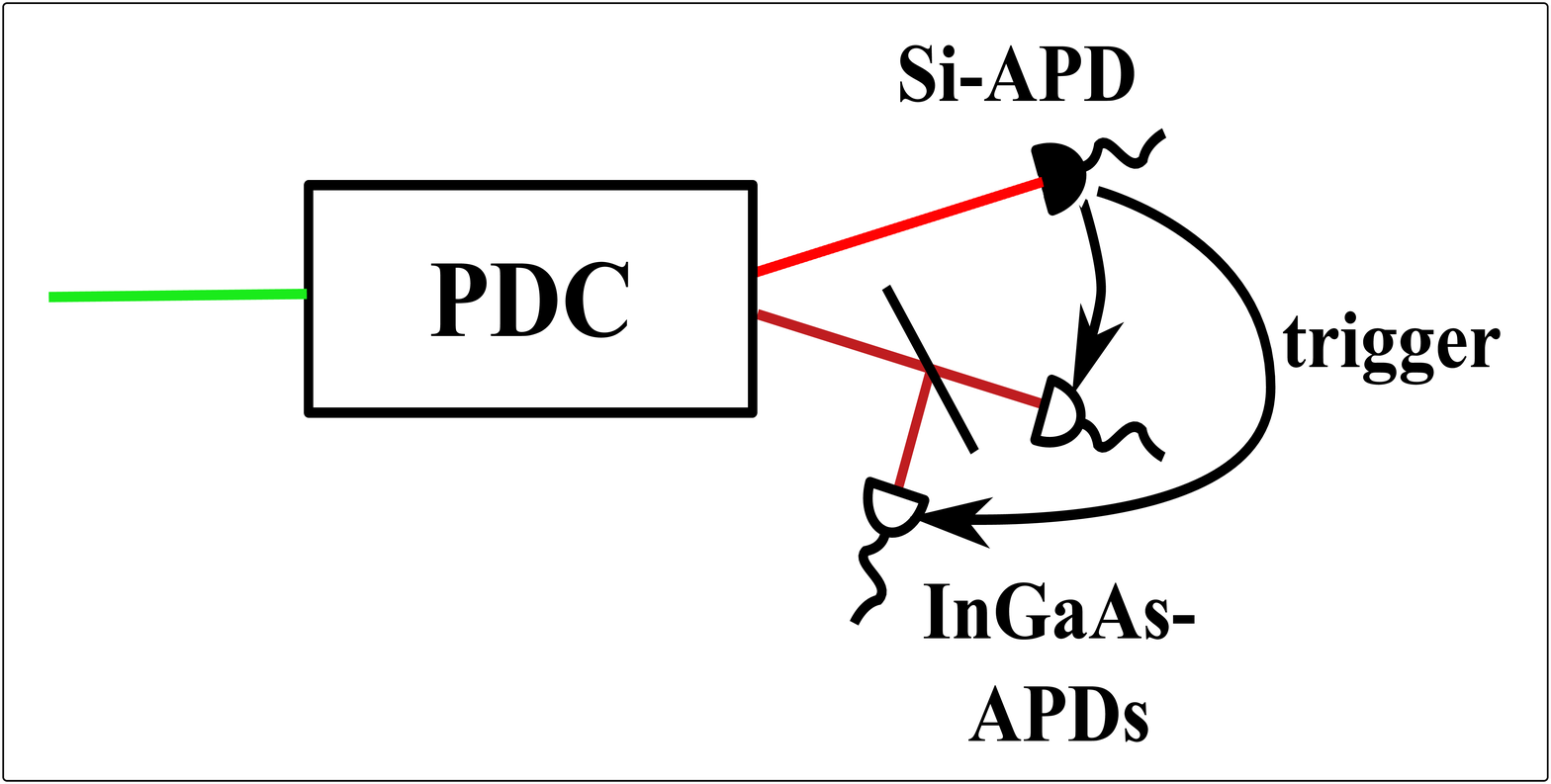}
\includegraphics[width=0.49\linewidth]{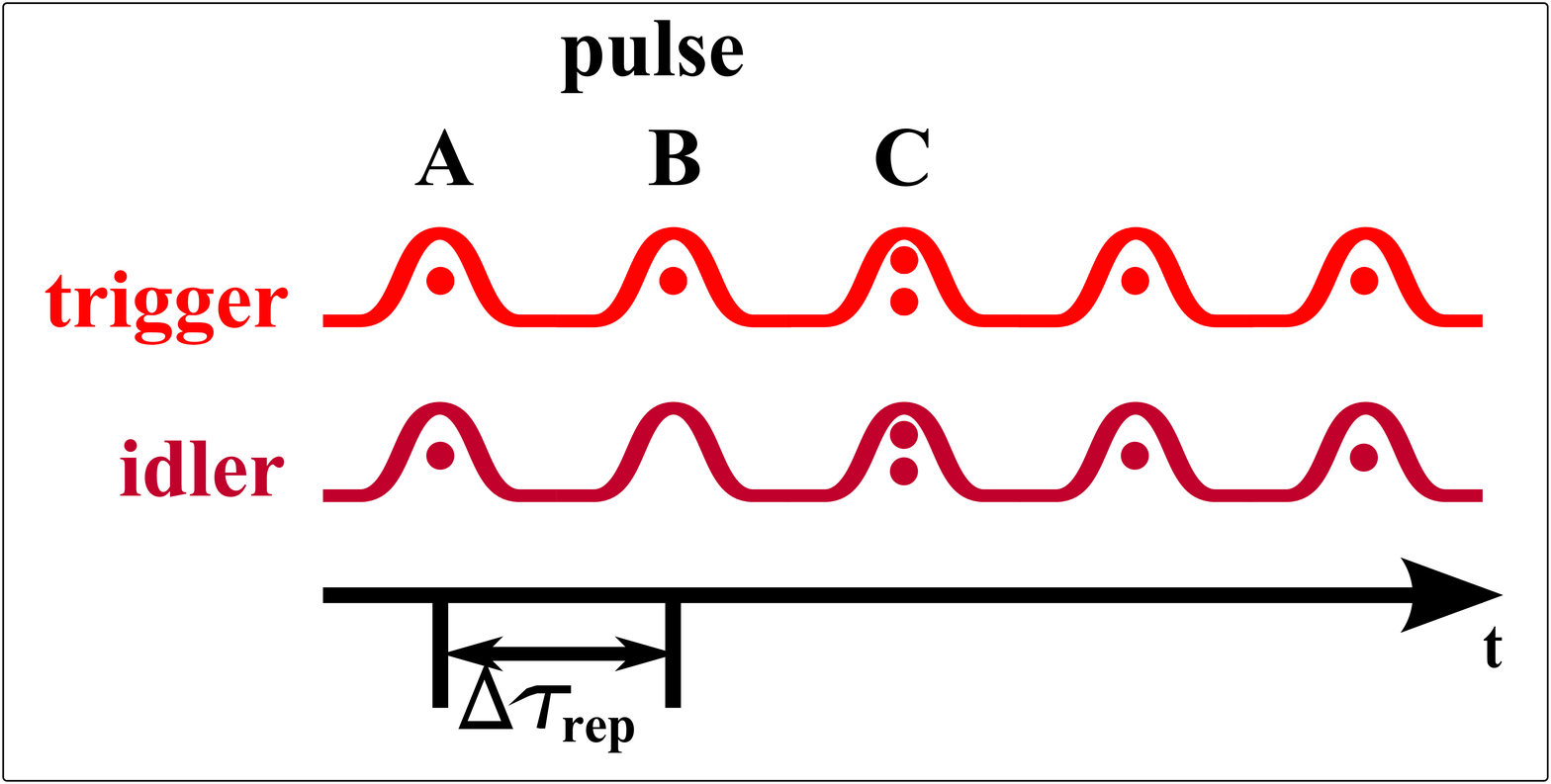}
\caption{\label{Pic2} Conditioned PDC detection scheme (left); Temporal PDC event behaviour with pulsed pump (right)}
\end{figure}

The scheme of our heralded single photon preparation is shown in \fref{Pic2} (left), where the Si-APD measures a detection rate $R_{\mathrm{Si}}$ with an efficiency $\eta_{\mathrm{Si}}$. The two InGaAs APDs are conditioned on a Si-APD detection event and register the rates of the collected idler photons $R_{\mathrm{Id,1}}$ and $R_{\mathrm{Id,2}}$. We assume similar detection efficiencies $\eta_{\mathrm{Id1}}=\eta_{\mathrm{Id2}}=\eta_{\mathrm{Id}}$ for both InGaAs detectors and neglect the influence of dark counts. Since the idler photons are sent through a $50/50$ beamsplitter in front of the detection, the rates $R_{\mathrm{Id,1}}$ and $R_{\mathrm{Id,2}}$ individually reflect genuine photon pair generation events, whereas three-fold coincidences between all detectors indicate the generation of at least two photon pairs in the PDC process. These occur with the rate $R_{\mathrm{c}}$. In this setting, the anti-correlation parameter $\alpha$ is given by \cite{Grangier1986,URen2005}
\begin{equation}
\label{7}
\alpha=\frac{R_{\mathrm{Si}}\cdot R_{\mathrm{c}}}{R_{\mathrm{Id,1}}\cdot R_{\mathrm{Id,2}}}.
\end{equation}

For judging the amount of higher order photon components in the generated PDC state, we can utilize the second-order auto-correlation or Glauber function $g^{(2)}(0)$. For our measurement scheme, it has been shown for example in \cite{Fasel2004}, that $g^{(2)}(0)$ is given by
\begin{equation}
\label{8}
g^{(2)}(0)\simeq4\frac{R_{\mathrm{Si}}\cdot R_{\mathrm{c}}}{\left(R_{\mathrm{Id,1}}+R_{\mathrm{Id,2}}\right)^2}
\end{equation}
if we assume that the probability for the generation of higher order components is much smaller than for the generation of single photon pairs. Note that, in the case discussed here, the two values $\alpha$ and $g^{(2)}(0)$ are both measures for the ratio between photon pairs and unwanted higher order components, although they are derived from different conceptual approaches.

The Klyshko efficiency \cite{Klyshko1980} $\eta_{\mathrm{K}}$ of the idler arm is given by
\begin{equation}
\label{9}
\eta_{\mathrm{K}}=\frac{R_{\mathrm{Id,1}}+R_{\mathrm{Id,2}}}{R_{\mathrm{Si}}}.
\end{equation}
and is a general measure for the generation of of single photon pairs. Since our InGaAs-APDs have efficiencies below one, we also calculate the heralding efficiency
\begin{equation}
\label{14}
\eta_{\mathrm{H}}=\frac{\eta_{\mathrm{K}}}{\eta_{\mathrm{Id}}}.
\end{equation}
which describes the performance of the single photon preparation in our experimental setup.

Like in continuous wave experiments, the coincidences-to-accidentals ratio (CAR) provides valuable information about the performance of our PDC source, but the fact that we work with pulsed light requires a careful definition of CAR in our experiment. \Fref{Pic2} (right) schematically illustrates typical measurement results. From this we immediately see, that the CAR is dependent on the relative time delay between signal and idler detection.

When shifting the delay of the trigger pulse $A$ and the measured correlated idler gradually to a point between two idler pulses, we expect decreasing Klyshko and heralding efficiencies. Relating the maximum Klyshko efficiency (i.e. perfect pulse overlap) to the minimum Klyshko efficiency, we get
\begin{equation}
\label{16}
CAR_{\mathrm{\Delta\tau}}=\frac{\eta_{\mathrm{K}}(P_{\mathrm{p}},\Delta\tau=0)}{\eta_{\mathrm{K}}(P_{\mathrm{p}},\Delta\tau=T)},
\end{equation}
where $|T|$ must be larger than the gating time window width but smaller than the repetition time $\tau_{\mathrm{rep}}$ of our pump. Whenever parasitic fluorescence with long lifetimes is induced, $CAR_{\mathrm{\Delta\tau}}$ will drop significantly due to artificial coincidences outside the region of perfect temporal pulse overlap. Note that our definition of $CAR_{\mathrm{\Delta\tau}}$ is equivalent to the inverse output noise factor (ONF), defined in \cite{Brida2011} and used for heralded single photon sources working in the cw regime \cite{Brida2012}.

By defining a different $CAR_{\mathrm{rep}}$, we indirectly get information about two-photon-pair contributions, if we shift the relative time delay such, that pulse $A$ overlaps with the neighbouring pulse $B$, given by an integer multiple of the repetition time $\tau_{\mathrm{rep}}$. Then, detected coincidences originate from non-correlated PDC events. From the expression
\begin{equation}
\label{17}
CAR_{\mathrm{rep}}=\frac{R_{\mathrm{c,double}}(P_{\mathrm{p}},\Delta\tau=0)}{R_{\mathrm{c,double}}(P_{\mathrm{p}},\Delta\tau=m\cdot\tau_{\mathrm{rep}})}, \;\;m\in\mathds{N}.
\end{equation}
we can deduce the photon pair generation probability $p$, which is closely linked to the mean photon numver per pulse and hence to the brightness of our source. The value of $CAR_{\mathrm{rep}}$ must be the same, regardless of the number of skipped pulses.

A third CAR evolves as a general measure of higher order photon contributions, if we relate the rate of all two-fold coincidences $R_{\mathrm{c,double}}$ to the rate of unwanted three-fold coincidences $R_{\mathrm{c}}$ at perfect temporal pulse overlap (i.e. $\Delta\tau=0$). The purity of our source - in addition to the $g^{(2)}(0)$ value - is then characterized by
\begin{equation}
\label{18}
CAR_{\mathrm{HOP}}=\frac{R_{\mathrm{c,double}}(P_{\mathrm{p}},\Delta\tau=0)}{R_{\mathrm{c}}(P_{\mathrm{p}},\Delta\tau=0)}.
\end{equation}
The value of $CAR_{\mathrm{HOP}}$ is expected to drop drastically with increasing pump powers according to the increasing generation of higher order photon pairs. Note that, for the case of perfect detectors, $CAR_{\mathrm{rep}}$ and $CAR_{\mathrm{HOP}}$ result in exactly the same values, whereas deviations of this equality can be solely contributed to imperfect detectors.

\subsection{Second building block - WDM coupler}
\label{Sec23}

In order to separate signal and idler photons spatially, we implemented a passive demultiplexer (WDM-coupler) behind the periodically poled area. 

\begin{figure}
\includegraphics[width=0.6\linewidth]{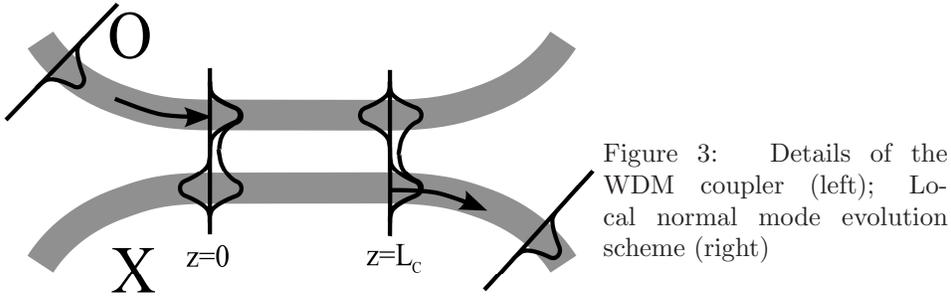}
\begin{minipage}[b]{0.35\linewidth}
\caption{\label{Pic3} Details of the WDM coupler (left); Local normal mode evolution scheme (right)}
\end{minipage}
\end{figure}

The most accurate way to treat a waveguide directional coupler is to describe its properties in terms of local normal modes of the dual channel structure \cite{Marcatili1969,Burns1976,Bersiner1991}. In our case the local normal modes comprise of a symmetric (00) and an anti-symmetric (10) mode, as shown in \fref{Pic3}. For sufficiently large channel gaps $d$ compared to the mode size, the propagation constants of the two modes are degenerate.

The launching of light into one input channel $O$ is described by constructive interference of these local normal modes in the launching channel and a destructive interference in the cross channel $X$. Upon approaching of the two channels towards the homogeneous central section of the coupler, the degeneracy of the propagation constants of the two modes is gradually removed, as the propagation constants exponentially depend on the center-to-center distance $d(z)$. This causes an increasing amount of phase difference 
\begin{equation}
\label{20}
\Delta\phi=\int_{z_1}^{z_2}\left(\beta_{00}(d(z))-\beta_{10}(d(z))\right)dz\ne0.
\end{equation}
and the corresponding interference change between the two modes. As a result the power flow gradually bounces from the launching channel to the adjacent one. Complete overcoupling to the cross channel is achieved when the accumulated phase difference is $\Delta\phi=\pi$. From this we can deduce, that the power flux will oscillate with a squared sinusoidal behaviour dependent on the coupler stem length $L_\mathrm{C}$.

Applying FEM calculations \cite{Bersiner1991,Strake1988}, we anticipated the respective propagation parameters very precisely. For our non-degenerate PDC process we designed the coupler such that first-order coupling is only provided for the fundamental waveguide mode at the idler wavelength whereas the signal power almost fully remains in the original channel due to the stronger confinement at the shorter signal wavelength. Bending losses are circumvented by the implementation of S-bend input and output ports with sufficiently large radii of curvature \cite{Tsao2002}. On our chip we only varied the coupler stem length from $2750\,\mu\mathrm{m}\le L_\mathrm{C}\le4250 \,\mu$m at a fixed center-to-center gap of $13\,\mu$m within different test structures to determine the optimum coupling behaviour of the idler wavelength, which is shown schematically in \fref{Pic3}. Fixing the output port gap to a value of $165\,\mu$m provides us with the opportunity of a direct fiber-connection via silicon V-groves.

As a measure for the demultiplexing dependence on the coupler's stem length $L_\mathrm{C}$, we define the power suppression in the unwanted output ports using a logarithmic scaling, where
\begin{equation}
S_{\mathrm{s}}=10\cdot \mathrm{log}\left(\frac{P_{\mathrm{X}}(L_{\mathrm{C}},\lambda_{\mathrm{s}})}{P_{\mathrm{O}}(L_{\mathrm{C}},\lambda_{\mathrm{s}})}\right)
\label{25}
\end{equation}
denotes the signal suppression in the cross port(X) and
\begin{equation}
S_{\mathrm{i}}=10\cdot \mathrm{log}\left(\frac{P_{\mathrm{O}}(L_{\mathrm{C}},\lambda_{\mathrm{i}})}{P_{\mathrm{X}}(L_{\mathrm{C}},\lambda_{\mathrm{i}})}\right)
\label{26}
\end{equation}
represents the idler suppression in the original port (O).

\begin{figure}
\includegraphics[width=0.60\linewidth]{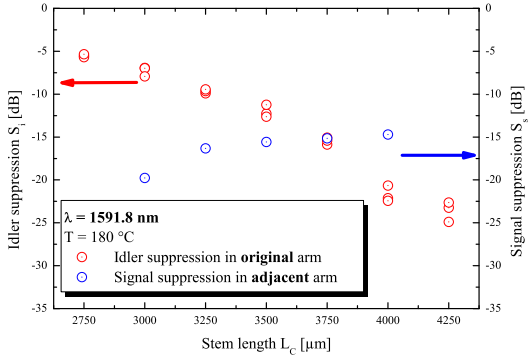}
\begin{minipage}[b]{0.35\linewidth}
\caption{\label{Pic6} WDM coupler characteristics for signal (red) and idler (blue)}
\end{minipage}
\end{figure}

We carefully analyzed the coupling properties using classical light. It turned out that the power suppression in the unwanted WDM output ports does depend on the stem length $L_\mathrm{C}$ as expected (see \fref{Pic6}). In the optimum case at $L_\mathrm{C}=4000\,\mu$m we get close-to-perfect coupling behaviour for both, signal and idler. While the signal remains with a ratio of $\eta_{\mathrm{WDM,s}}=96.5$\% in the original channel, the idler couples by $\eta_{\mathrm{WDM,i}}=99.1$\% to the cross channel, corresponding to suppressions of $S_{\mathrm{s,opt}}=-15$ dB and $S_{\mathrm{i,opt}}=-20.6$ dB, respectively.

\subsection{Waveguide, WDM and poling technology}
\label{Sec24}
The realization of low loss waveguides and WDM coupler structures for efficient type-I PDC has been achieved using Z-cut lithium niobate propagation along X-direction.

In our specific case, titanium of $80$ nm homogeneous thickness was deposited by electron beam evaporation onto the Z-surface of the substrate. The waveguide structure was defined by photolithograpic delineation of the titanium layer using deep UV contact printing. The residual Ti stripes were indiffused at $1060^{\circ}$C for $8.5$ hours in an oxygen-rich atmosphere, allowing for the idler wave to be guided in its fundamental spatial mode, whereas pump and signal photons can be guided in multiple modes.

\begin{figure}
\includegraphics[width=0.48\linewidth]{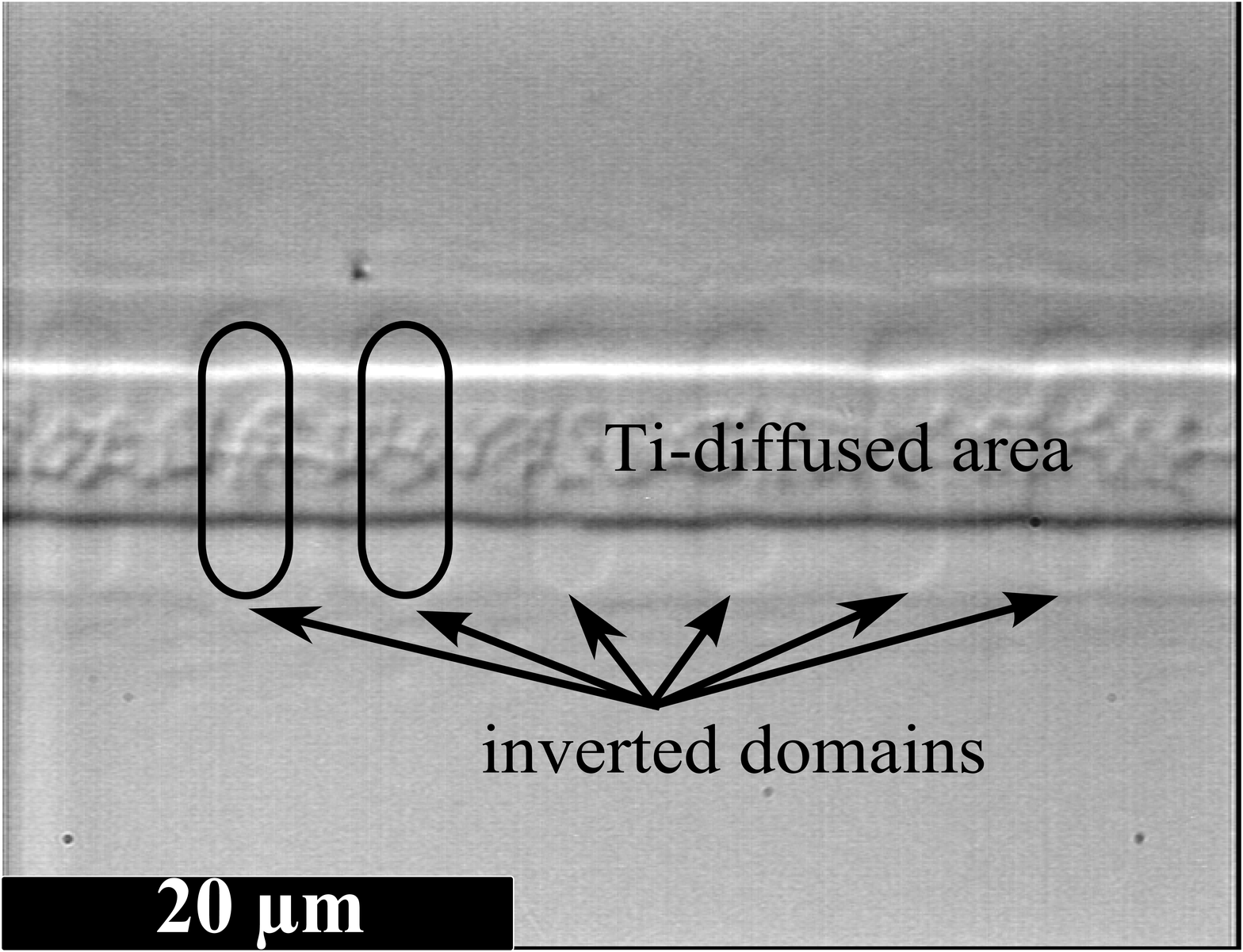}
\includegraphics[width=0.51\linewidth]{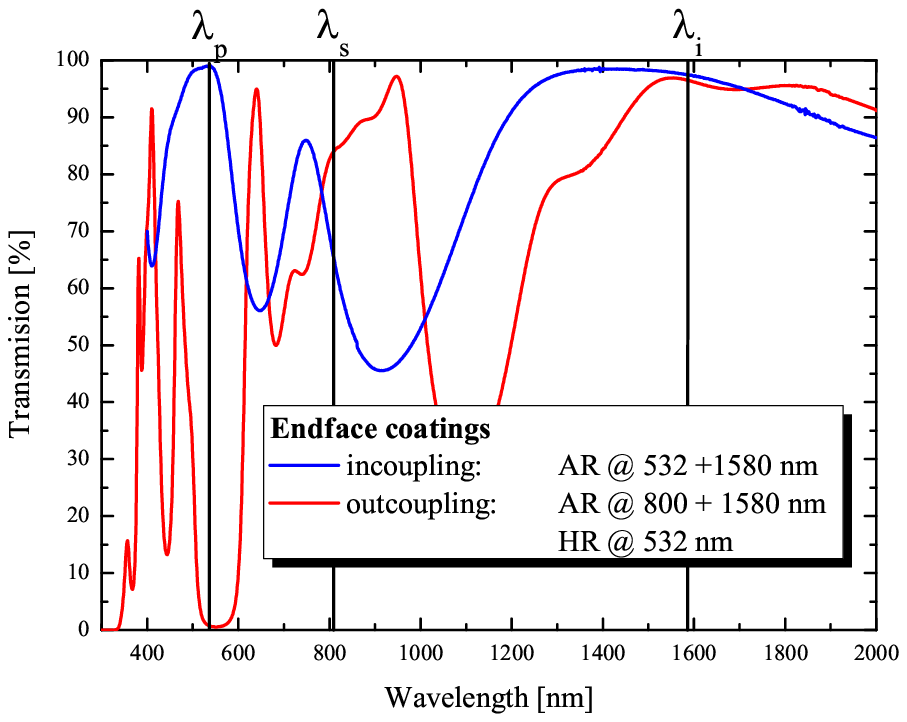}
\caption{\label{Pic4} Domain structure of the periodically poled waveguide area (left); characteristics of the end-face coatings (right)}
\end{figure}

For periodic poling the standard electric field assisted poling method \cite{Yamada1993} was applied, using a waveguide selective segmented resist patterning for electrical insulation and lithium chloride solution as the contact electrolyte. By applying short voltage pulses of 11.5 kV and monitoring the displacement current due to the periodic polarization reversal, the accumulated charge was controlled to provide the proper duty cycle of the $30$ mm long domain grating. A typical poling structure is visualized in \fref{Pic4} (left).

In order to achieve the best possible coupling of light into and out of the waveguides, the end-facettes were polished perpendicularly to the propagation direction. We designed and deposited different end-face coatings (EFC) for optimized incoupling of the pump as well as for the best possible outcoupling of signal and idler wavelengths including the reflection of the pump. The transmission characteristics for both coatings are shown in \fref{Pic4} (right). Note that the coating at the waveguide output reaches more than $99\%$ of reflection for the pump, while the idler wavelength is transmitted with more than $96.5\%$.

An important linear optical feature of the waveguides is their low intrinsic loss of $0.07$ dB/cm in average, which we measured for straight control channels next to the coupler structures. For this measurement we applied the well-established low-finesse Fabry-P\'erot-interferometric method \cite{Regener1985}.

An issue for later nonlinear measurements arises from the fact, that the mode size of the idler wave and its symmetry both are important for the coupling to single mode fibers (SMF), since any asymmetry decreases the overlap. Therefore we measured the waveguide mode sizes at the idler wavelength and calculated a theoretical mode overlap integral to standard single mode fibers in TM polarization of $90.1$\%, which is the benchmark for the Klyshko efficiency under the assumption of perfect transmission in the idler beam path and perfect detectors.

\section{Experimental Setup}
\label{Sec3}

\begin{figure}
\includegraphics[width=0.99\linewidth]{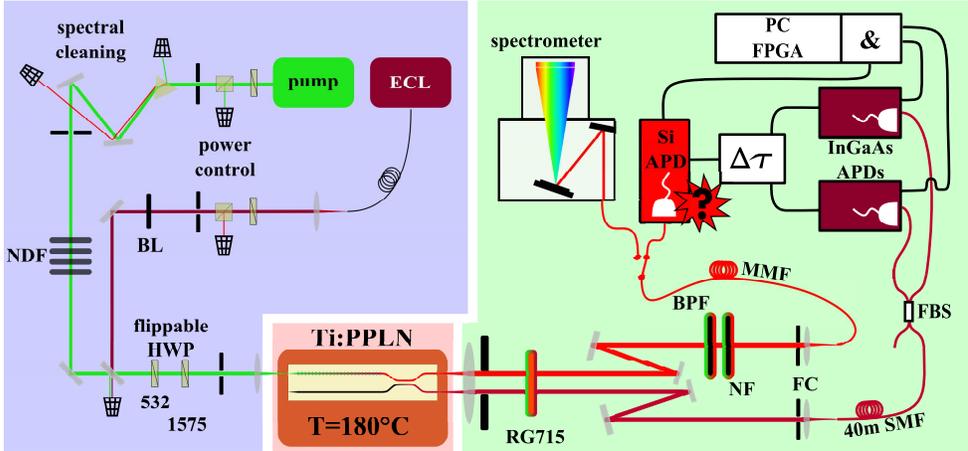}
\caption{\label{Pic5} Experimental setup for spectral characterization and conditioned measurements; NDF: neutral density filters, BL: beam blocker, ECL: external cavity laser for idler coupling adjustment, HWP: half-wave plate, RG715: customized absorptive filter, NF: needle filter, BPF: band-pass filter, FC: fiber coupling stages, SMF: single mode fiber, MMF: multi-mode fiber, FBS: fiber beamsplitter $50/50$, $\Delta\tau$: delay generator, APD: avalanche photodiode, FPGA: field programmable gate array}
\end{figure}

\Fref{Pic5} illustrates our experimental setup, which can be separated into three major divisions: incoupling, sample adjustment and analysis part.

We used a frequency-doubled pump source (Katana-05, Onefive GmbH Z\"urich), which converts amplified diode laser light at $1064$ nm to the green in a second harmonic generation process. The laser is running at $10$ MHz repetition rate with Fourier-limited pulses of $43$ ps duration and Gaussian shape. Its spectral width is specified with $0.19$ nm while it offers a low pulse-to-pulse timing-jitter of less than $10$ ps. In order to spectrally filter out any residual $1064$ nm pump light we added a prism-based cleaning stage in front of a variable attenuator and neutral density filters, which allow for additional power control. The required TM polarization for the type-I PDC process is set using a $\lambda/2$-plate right in front of an AR-coated incoupling lens. 

The PDC source is stabilized at $T_{\mathrm{OP}}=(185\pm0.2)^{\circ}\mathrm{C}$ on a home-assembled 5-axis stage with high precision. The analysis part of the setup is kept in a free-space configuration as far as possible. 

Right behind the AR-coated outcoupling lens we placed a home-deposited RG$715$ filter, which transmits signal and idler with $97$ \% and has an overall pump suppression larger than six orders of magnitude.

Two D-shaped mirrors further split up the on-chip spatially separated beams to a signal and an idler arm. The signal arm consists of a multimode fiber coupling stage, which is connected to a spectrometer system (Andor iKon-M $934$P-DD/Shamrock SR-$303$iA) for spectral PDC characterization. It can also be connected to the free-running silicon avalanche photo diode (Si-APD, Perkin Elmer SPCM-AQRH-13) for the heralding experiments and for conditioned $g^{(2)}(0)$ measurements at room temperature. The Si-APD is specified to have a detection efficiency of $\eta_{\mathrm{Si}}(\lambda_{\mathrm{s}})=55\%$ at around $800$ nm and was measured to have an average dark count rate of $R_{\mathrm{T,dark}}=(238\pm16)\,\mathrm{s}^{-1}$. To extract the PDC signal photons from any noisy background in the heralding experiments, we can flip an additional bandpass-filter ($FWHM=12$ nm, $T_{\mathrm{peak}}\ge99.5$\%) as well as a needle filter ($FWHM=0.5$ nm, $T_{\mathrm{peak}}=78$\%) into the signal beam path.

The idler arm also includes a high precision fiber coupling stage for injecting the idler photons to the InGaAs-APDs (ID Quantique ID201) via $40$ m of AR-coated single mode fiber, the length of which ensures the optical compensation of electrical delays. For $g^{(2)}(0)$ measurements, a fiber beam splitter enables us to address a second InGaAs-APD. At the idler wavelength both APDs have specified detection efficiencies of $\eta_{\mathrm{Id}}(\lambda_i)=0.23$. Their aquisition time gate for idler click events was set to $2.5$ ns with zero additional deadtime.

Electronical compensation of the occuring temporal gap in the arrival of signal and idler at the respective detector is realized using a Stanford Research DG645 delay generator. Data aquisition and the evaluation of single and coincidence events are carried out using an FPGA interface card operating at $40\,\mathrm{MHz}$.

A classical characterization of the overall idler transmission loss from the device's endface to the InGaAs-APDs input resulted in a transmission factor of $T_i=0.663$. This includes the device's endface throughput as well as filters, lenses, fiber coupling and the optical loss in the fiber beamsplitter.

In contrast to the benchmarking Klyshko efficiency described in \sref{Sec24}, the maximum Klyshko efficiency of our experimental setup is limited by the overall transmission in the idler arm, by the respective detector efficiency and by the coupling efficiency of the on-chip integrated WDM coupler to
\begin{equation*}
\eta_{\mathrm{K,System}}=T_i\cdot\eta_{\mathrm{Id}}\left( \lambda \right)\cdot\eta_{\mathrm{WDM,i}}=0.151
\end{equation*}
resulting in a maximum achievable heralding efficiency of $\eta_\mathrm{H,max}=65.6$\%. This upper experimental boundary only holds for very low pump power levels, where higher order photon pair contributions can be neglected.

\section{Results and Discussion}
\label{Sec4}

\subsection{Spectral characteristics}
\label{Sec41}

\begin{figure}
\includegraphics[width=0.49\linewidth]{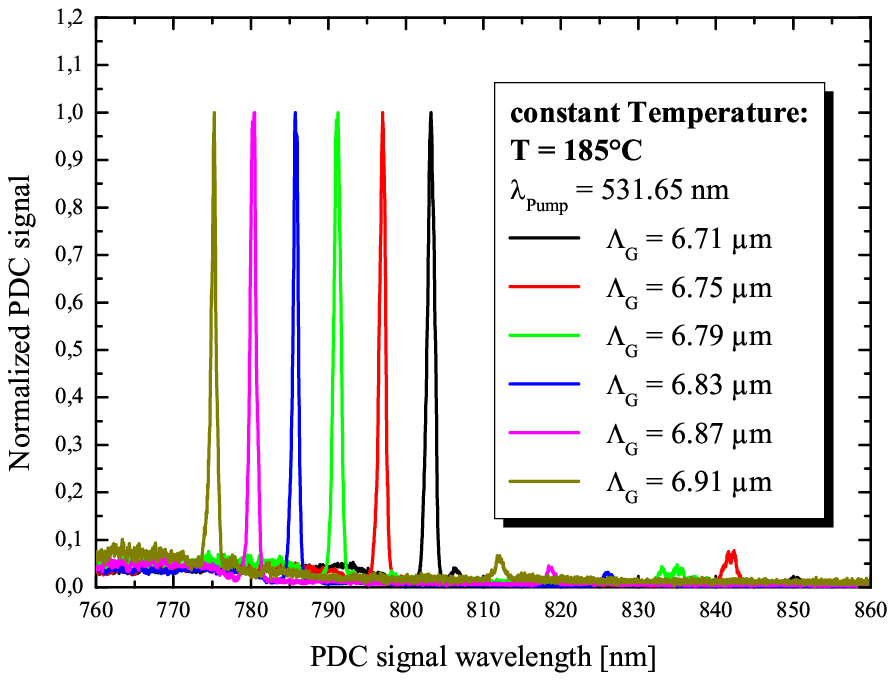}
\includegraphics[width=0.49\linewidth]{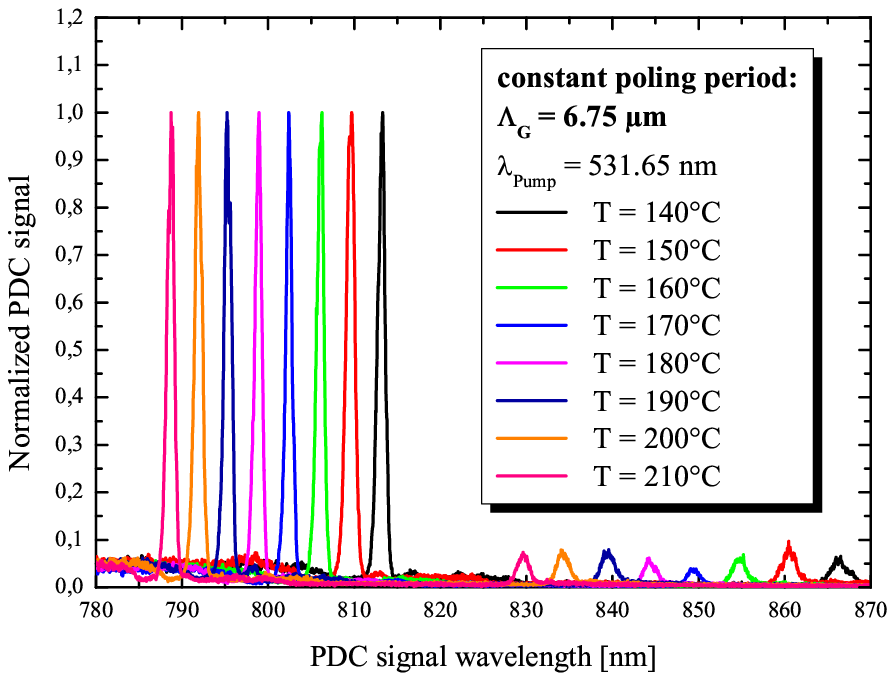}
\caption{\label{Pic7} Tunability of the PDC signal wavelength with the crystal's poling period (left) and the device's temperature (right)}
\end{figure}

The spectral dependencies of the generated PDC signal on temperature and poling period have been determined and are plotted as normalized curves in \fref{Pic7}. These results prove a convenient wavelength tunability to be either coarse (with poling period) or fine (with temperature). Note that there are practically no significant sidepeaks or higher order mode excitations indicating an excellent waveguide homogeneity and the best possible incoupling of the pump to its fundamental waveguide mode. In any case the signal FWHM is around $0.7$ nm, which is mainly contributed by the on-chip nonlinear interaction length of $30$ mm, while the influence of the spectral width or limiting temporal properties of the pump are neglectable. We can clearly identify a broad structure in the lower wavelength region next to the PDC signal peaks in \fref{Pic7}. We suspect this to be nonlinear $\mathrm{\check{C}}$erenkov radiation associated with the quasi-phasematching process \cite{Zhang2008a}. These background effects have been suppressed for the following experiments with the help of a needle filter as mentioned in \sref{Sec3}.

\subsection{State preparation rates}
\label{Sec42}

\begin{figure}
\includegraphics[width=0.60\linewidth]{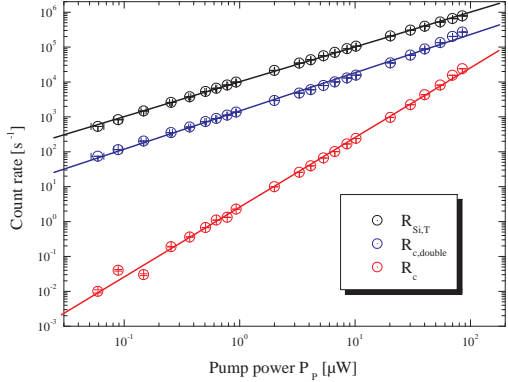}
\begin{minipage}[b]{0.35\linewidth}
\caption{\label{Pic9} Pump-dependent count rates of the trigger (black), the sum of the two-fold coincidences (blue)  and the triple coincidences (red) in conditioned measurements. The solid lines serve as a guide to the eye.}
\end{minipage}
\end{figure}

Our power-dependent $g^{(2)}(0)$-measurements yielded the trigger rates $R_{\mathrm{Si}}$, the overall two-fold coincidence rates $R_{\mathrm{c,double}}=R_{\mathrm{Id,1}}+R_{\mathrm{Id,2}}$ and the triple coincidence rates $R_{\mathrm{c}}$ within $1$ s of measurement time. Repeating this hundredfold provides us with small error bars. The three-fold coincidence rate contributes the highest uncertainities. \Fref{Pic9} shows the respective rates in double-logarithmic scaling. The trigger rate has an almost linear dependency on the increasing pump power, while the coincidence rates both increase superlinearly in the pump power range above $10\,\mu$W. Note that the linearity of the trigger rate disappears above $10^{5}$ counts per second due to deadtime effects.

\subsection{Heralded single photon preparation}
\label{Sec43}

\begin{figure}
\hspace{0.5mm}
\includegraphics[width=0.49\linewidth]{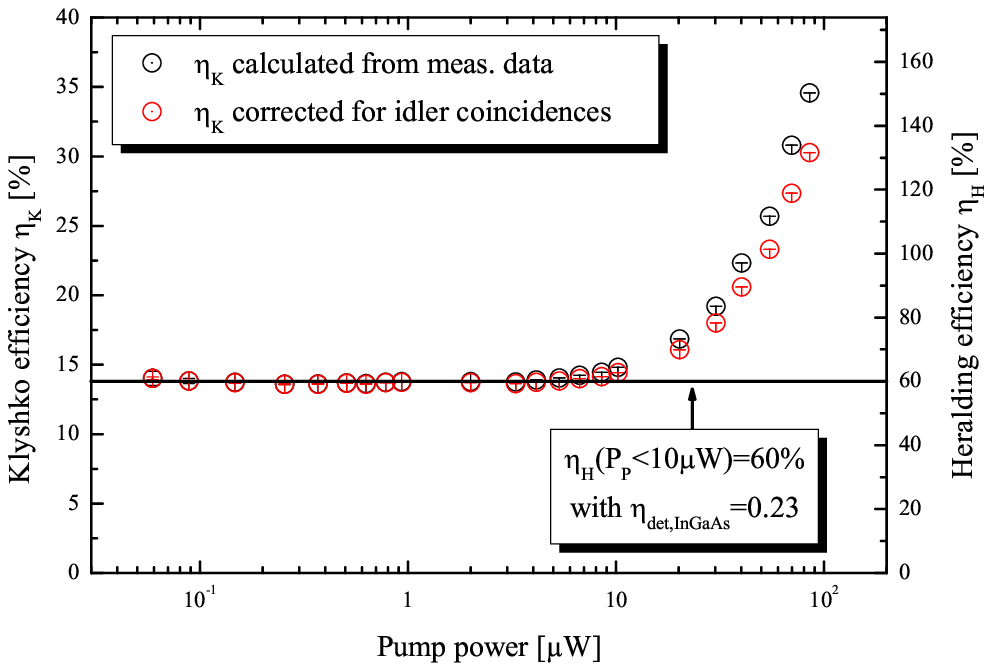}
\includegraphics[width=0.49\linewidth]{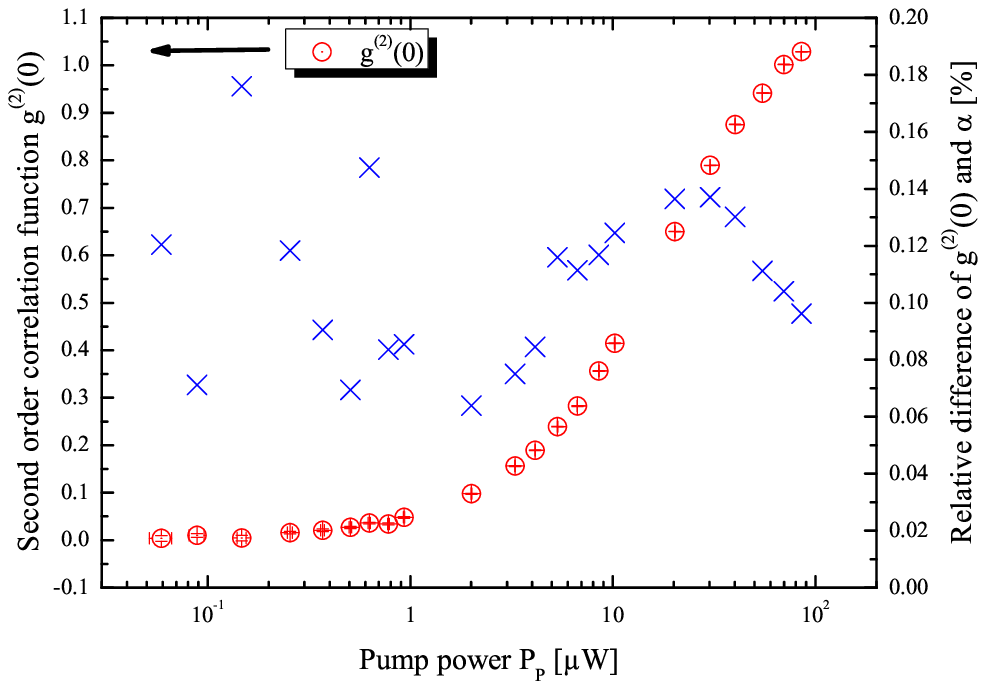}
\caption{\label{Pic10} Pump-dependent Klyshko efficiency and heralding efficiency (left); auto-correlation function $g^{(2)}(0)$ and its relative difference to the anti-correlation parameter $\alpha$ (right)}
\end{figure}

From the count rates we calculated the Klyshko and heralding efficiencies according to \eref{9} and \eref{14}, which are plotted in \fref{Pic10} (left). The $g^{(2)}(0)$ values and the anti-correlation parameter $\alpha$ were calculated using \eref{8} and \eref{7}, respectively. 

We identified an almost constant heralding efficiency as high as $60\%$ at pump powers less than $10\,\mu$W, where seemingly only first order photon pairs contribute to the measurement (compare plots in \fref{Pic9}) and the trigger rate is up to $R_{\mathrm{Si}}=105$ kHz. To the best of our knowledge, this is the first time, that in a Ti:PPLN based type-I PDC source such high heralding efficiencies have been reported, while other sources using comparable detection schemes (see for example \cite{Mosley2008,Soller2011a,Steinlechner2012}) rely on bulk nonlinear materials or photonic crystal fibers and do not offer more than one integrated functonality. The coupling limitations of our lab system can probably be overcome, when a completely fiber-based setup is used in the future. This should further improve the heralding efficiencies to around $80$\%.

As shown in \fref{Pic10} (right), the auto-correlation function at very low power levels ($P_{\mathrm{p}}\le1\mu$W) reaches a minimum value of $g^{(2)}(0)=0.0038$, which is one of the best results for heralded single photon sources in general, while it marks the lowest obtained value in titanum-indiffused PPLN waveguide PDC sources to date. It could be driven even lower at the cost of the heralding rate. But for applications like QKD with this kind of PDC sources, low trigger rates are always related to low key transmission rates, and a trade-off between high rates and purity of the single photons is unavoidable. 

From our results we can conclude, that the approximation made in \eref{8} and in \cite{URen2005,Fasel2004} is only valid and suitable for low pump powers. At the upper end of our power scale, we see that $g^{(2)}(0)$ approaches the value of one, which would correspond to an infinite average photon number. Note however, that at these high pump power levels our assumption, that the probability of generating higher order photon pairs being much smaller than the probability of single photon pairs, remains no longer valid. The triggering rate $R_{\mathrm{Si}}$ as well as the triple coincidence rate $R_{\mathrm{c}}$, both tend to be overestimated due to higher order photon contributions. The Si-APD fires only once, when there is more than one trigger photon arriving and internal deadtime effects of the InGaAs-APDs limit the repetition rate of the detection of double coincidences. We can derive from the graph in \fref{Pic9} in conjunction with \fref{Pic10} (right), that trigger rates higher than $310\mathrm{kHz}$ should be avoided for this kind of detectors in combination with our kind of measurement. It also shows, that the relative differences between the calculated values of $g^{(2)}(0)$ and the anti-correlation parameter $\alpha$ are very small, which is a proof of the equivalence of \eref{7} and \eref{8} for our measurements.

\subsection{Measurement of CARs}
\label{Sec44}

\begin{figure}
\includegraphics[width=0.60\linewidth]{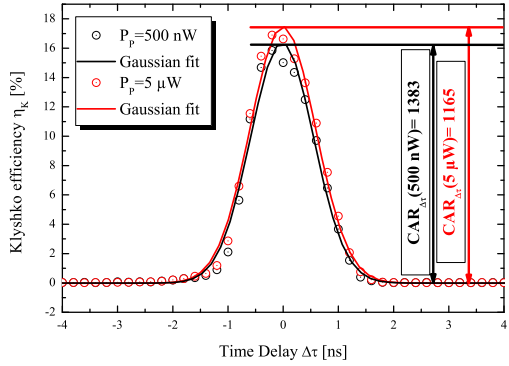}
\begin{minipage}[b]{0.35\linewidth}
\caption{\label{Pic11} Dependency of the Klyshko efficiency on the temporal overlap between signal and idler detection}
\end{minipage}
\end{figure}

\begin{figure}
\includegraphics[width=0.60\linewidth]{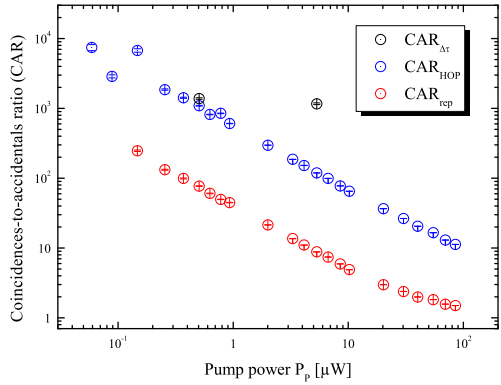}
\begin{minipage}[b]{0.35\linewidth}
\caption{\label{Pic12} Pump dependency of the three calculated coincidence-to-accidentals ratios}
\end{minipage}
\end{figure}

For the analysis of detrimental background photons we determined the source's heralding efficiency around the optimum time delay between trigger and heralded photons at two different power levels, measuring the count rates in step widths of $0.2$ ns for $5$ s, each. The characteristics in \fref{Pic11} show almost Gaussian shape. This is mainly caused by the intrinsic gate width of the InGaAs-APDs of $\tau_{\mathrm{gate}}=(1.16\pm 0.32)$ ns, where the error includes the timing jitter of the InGaAs-APD, the Si-APD and of the delay generator. The graph visualizes the slight increase of the Klyshko efficiency with increased pump power due to higher order photon pair contributions. In \fref{Pic12} we see, that the Klyshko efficiency is almost not affected by pump-induced fluorescence of any optical component. In particular, at $500$ nW and $5\,\mu$W the $CAR_{\mathrm{\Delta\tau}}$ have been calculated to be $1383$ and $1165$, respectively and thus do not decrease significantly. Please note, that in \fref{Pic11} this feature cannot to be seen intuitively due to the scaling of the accidentals. Inverting the $CAR_{\mathrm{\Delta\tau}}$ values provides us with output noise factors (ONF) of $0.072\%$ and $0.086\%$, respectively, which are both a factor of $3$ smaller than the best ONFs in comparable cw experiments \cite{Brida2012}. Together with the fact, that we are still in the single photon regime ($0.027\le g^{(2)}(0)\le0.24$) and with the high heralding rates $5.3\,\mathrm{kHz}\le R_{\mathrm{Si}}\le56\,\mathrm{kHz}$, this impressively points out the benefits of heralding single photon experiments in the pulsed regime.

\Fref{Pic12} shows that the $CAR_{\mathrm{rep}}$ for two adjacent pulses asymptotically tends to the value of $1$ at the highest suitable pump power, while the $CAR_{\mathrm{HOP}}$ dropped from $7440$ to around $10$ . This emphasizes the impact of higher order photon contributions and the usefulness of a trade-off between high heralding rates and a low $g^{(2)}(0)$.

\begin{figure}
\includegraphics[width=0.60\linewidth]{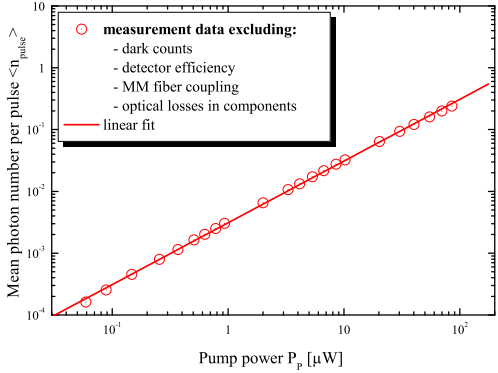}
\begin{minipage}[b]{0.35\linewidth}
\caption{\label{Pic13} Generated mean photon number per pulse in dependency on the pump power}
\end{minipage}
\end{figure}

One of the most important and remarkable figures of merit of our source is its brightness. We determined the pump dependent mean number of generated photons per pulse by dividing the heralding rate $R_{\mathrm{Si}}$ by the transmissivities of all optical components in the signal arm and by the repetition rate $R_{\mathrm{rep}}=10\,\mathrm{MHz}$, which is shown in \fref{Pic13}. We identified a maximum mean photon number per pulse of $\langle n_{\mathrm{pulse,max}}\rangle=0.24$ at the highest applied pump power of $85.14 \,\mu$W. A linear extrapolation yields an average photon number per pulse of $\langle n_{\mathrm{pulse,th}}\rangle\approx 34$ at $P_{\mathrm{P}}=10$ mW of cw-equivalent pump power, which is easily available with our pump laser.

\section{Conclusion and outlook}
\label{Sec5}
We demonstrated the on-chip integration of an efficient waveguide-based type-I PDC source in combination with a passive WDM coupler for spatial separation of signal and idler photons. It shows close-to-perfect demultiplexing behaviour at wavelengths of $803$ nm and $1575$ nm as well as very low scattering loss. We achieved high heralding efficiencies of $60\,\%$ in conjunction with a second order auto-correlation function as low as $0.0038$. At pump power levels below $10\,\mu$W only a minor influence of higher order photon pairs to the detection can be deduced from the results. We determined a coincidence-to-accidentals ratio of $CAR_{\mathrm{HOP}}>7400$ at low power levels and output noise factors $ONF<0.1$\% in the regime of low auto-correlation function values for our pulsed experiments. This indicates almost pure single photon pair generation together with high (although unoptimized) heralding rates as well as practically no influence of background noise. 

The high brightness of the source was derived from the extrapolation of the mean photon number per pulse of $\langle n_{\mathrm{pulse,th}} \rangle \approx 34$ at $P_{\mathrm{P}}=10$ mW cw-equivalent pump power.

At higher power levels $P_{\mathrm{P}}\ge30\,\mu$W our commercially available Si- and InGaAs-detectors tend to saturate, which distorts the calculation of correct $g^{(2)}(0)$ values, while the detection part of our analysis seems to be the limiting factor in terms of temporal resolution and dark counts. This limitation could be lifted in the future by novel detectors exhibiting drastically improved temporal properties (e.g. in \cite{Verma2012,Marsili2012}) in conjunction with a careful characterization of photon statistics using photon number resolving detection schemes (see \cite{Achilles2003,Fitch2003,Mauerer2007}). Then pump lasers with higher clock rates could yield higher heralding rates, which would be required for the commercialization, e.g., of quantum cryptography applications.

Besides that, more difficulties can arise in the used heralding scheme regarding the transmission of timing information within a wide area quantum network (compare \cite{Giustina2012}). Note however that - in contrast to cw realizations - in our pulsed scheme an intrinsic clock can be transferred from sender to receiver for the security of commercial QKD systems (compare, for example, \cite{Lydersen2010}).

Due to the strong dependence of the heralding efficiency on the coupling of the heralded photons to optical fibers, we will proceed in our work with the implementation of an all-fiber-connected device. This could increase the heralding efficiency to around $80\%$.

The general design of our device together with its figures of merit the constitute an important step for the implementation of integrated sources with multiple functionalities in future quantum networks with miniaturized nodes. Additional functionalities such as electro-optic phase shifters or polarization controllers in combination with passively working add-ons like polarization splitters and sophisticated endface-coatings could open new routes to establish quantum networks with novel node architectures.

\ack
The authors want to thank R. Ricken for the helpful technology discussions. This work was funded by the European Union in the QuRep project (reference: 247743).

\section*{References}
\bibliography{Literatur-Database}
\end{document}